%% file: main.tex
\documentclass[10pt]{IEEEtran}
 \usepackage{setspace}
\usepackage{xcolor} 
\usepackage{soul}
\soulregister\cite7
\soulregister\ref7
\soulregister\pageref7
\soulregister\ac7
\soulregister\acp7
\renewcommand{\hl}[1]{#1}

\usepackage{microtype}

\usepackage{amsmath}
\usepackage{amsfonts}
\usepackage{cite}
\usepackage{listings}
\usepackage{xcolor}
\usepackage{siunitx}
\usepackage{subfig}

\usepackage[english]{babel}
\usepackage{soul}

\usepackage[nolist,nohyperlinks]{acronym}
\usepackage[colorinlistoftodos]{todonotes}

\usepackage[hidelinks]{hyperref}

\usepackage{tikz}
\usetikzlibrary{arrows, petri, topaths}%
\usepackage{tkz-berge}

\lstdefinestyle{mystyle}{
    basicstyle=\ttfamily\footnotesize,
    breakatwhitespace=false,
    breaklines=true,
    captionpos=b,
    keepspaces=true,
    numbersep=5pt,
    showspaces=false,
    showstringspaces=false,
    showtabs=false,
    tabsize=2
}
\begin{document}

%\title{Symbolic congestion control policies from reinforcement learning experts}
\title{Closed-Form Congestion Control via Deep Symbolic Regression}

\author{
    \IEEEauthorblockN{ Jean Martins, Igor Almeida, Ricardo Souza, Silvia Lins }\\
    \IEEEauthorblockA{Ericsson Research, Indaiatuba, SP -- BR\\
        Email: \{jean.martins, igor.almeida, ricardo.s.souza, silvia.lins\}@ericsson.com
    }
}
% \orcid{0000-0003-2112-3723}

\newcommand{\Qval}[3]{Q_{#1}(#2, #3)}

\maketitle

\input{sections/abstract}
\input{sections/introduction}
\input{sections/background}

\input{sections/methodology}

\input{sections/results}

\input{sections/conclusions}

\begin{acronym}
    \acro{5G}{5th Generation Mobile Networks}
    \acro{A3C}{Asynchronous Advantage Actor Critic framework}
    \acro{AI}{Artificial Intelligence}
    \acro{B5G}{Beyond 5G}
    \acro{BBR}{Bottleneck Bandwidth and Round trip propagation time}
    \acro{BBU}{Baseband Unit}
    \acro{CC}{congestion control}
    \acro{CPRI}{Common Public Radio Interface}
    \acro{C-RAN}{Centralized Radio Access Networks}
    \acro{DCTCP}{Data Center TCP}
    \acro{DDPG}{Deep Deterministic Policy Gradient}
    \acro{DQN}{Deep Q-Network}
    \acro{DRL}{Deep Reinforcement Learning}
    \acro{DT}{Decision Tree}
    \acro{ECN}{Explicit Congestion Notification}
    \acro{FC}{Flow Control}
    \acro{GBT}{Gradient Boosting Trees}
    \acro{IETF}{Internet Engineering Task Force}
    \acro{IID}{Independent and Identically Distributed}
    \acro{MDP}{Markov Decision Process}
    \acrodefplural{MDP}{Markov Decision Processes}
    \acro{ML}{Machine Learning}
    \acro{NN}{Neural Network}
    \acro{NS-3}{Network Simulator 3}
    \acro{OOD}{Out of Distribution}
    \acro{PAL}{Partial Action Learning}
    \acro{RL}{Reinforcement Learning}
    \acro{RTT}{Round Trip Time}
    \acro{RRU}{Remote Radio Unit}
    \acro{RNN}{Recurrent Neural Network}
    \acrodefplural{RNN}{Recurrent Neural Networks}
    \acro{SL}{Supervised Learning}
    \acro{TCP}{Transmission Control Protocol}
    \acro{TD3}{Twin-Delayed Deep Deterministic Policy Gradient}
    \acro{TD}{Temporal Difference}
\end{acronym}

\bibliographystyle{IEEEtran}
\bibliography{references}

\end{document}

%% file: sections/abstract.tex
%removing page numnbers as indicated by globecom submission guidelines
\pagenumbering{gobble}

 \begin{abstract}

As mobile networks embrace the 5G era, the interest in adopting Reinforcement Learning (RL) algorithms to handle challenges in ultra-low-latency and high throughput scenarios increases. Simultaneously, the advent of packetized fronthaul networks imposes demanding requirements that traditional congestion control mechanisms cannot accomplish, highlighting the potential of RL-based congestion control algorithms. Although it is feasible to learn RL policies optimized for satisfying the stringent fronthaul requirements, neural network models' adoption in real deployments still poses some challenges regarding real-time inference and interpretability. This paper proposes a methodology to deal with such challenges while maintaining the performance and generalization capabilities provided by a baseline RL policy. The method consists of (1) training a congestion control policy specialized in fronthaul-like networks via reinforcement learning, (2) collecting state-action experiences from the baseline, and (3) performing deep symbolic regression on the collected dataset. The proposed process overcomes the challenges related to inference-time limitations through closed-form expressions that approximate the baseline performance (link utilization, delay, and fairness), and which can be directly implemented in any programming language. Finally, we provide an analysis of the closed-form expressions' inner workings.


\end{abstract}

%% file: sections/introduction.tex
\section{Introduction}
\label{sec:introduction}

Emerging \ac{5G} sparked the interest in more flexible, adaptable, 
and cost-efficient network architectures. Playing a fundamental role in this context, \ac{C-RAN}~\cite{jaber20165g} 
offer flexibility and reduced deployment costs, splitting the
regular radio base stations into the Baseband Unit (BBU) and the Remote Radio Unit (RRU). With the 
reduction of RRU node complexity comes the increasing throughput demands imposed on the fronthaul links, 
50 times higher than backhaul in some cases~\cite{holma2012lte}.

%\begin{figure}[!hb]
  %\centering
    %\includegraphics[width=.45\textwidth]{figs/C-RAN_dia}
    %\caption{C-RAN functional split showing centralized Baseband Units and remote Radio Units.}%
    %\label{fig:cran-split1}
%\end{figure}

Transitioning from dedicated fiber links (which adopted \ac{CPRI}~\cite{chitimalla20175g} protocol) 
towards packetized network deployments (i.e., with statistical multiplexing) is one strategy to reach more 
cost-efficient fronthaul solutions. However, shifting to a shared infrastructure introduces additional 
challenges for the transport network: fronthaul links may experience congestion due to, e.g., aggressive 
radio schedulers. Traditional \ac{CC} algorithms often face limitations when dealing with 
low latency and high throughput demands, leaving the challenge of addressing congestion control in packetized 
fronthaul networks.

There has been much exploration regarding using machine learning techniques for improvements in TCP \ac{CC} algorithms, 
with promising results~\hl{\cite{jay2019deep, zhang2020machine, congcontrol2021renaissance}}. More recently, the interest 
increased regarding the adoption of deep \ac{RL} algorithms that learn completely new \ac{CC} policies from  
scratch~\cite{xiao2019tcpdrinc, liu2019tyrus,abbasloo2020sigcomm,eagle2020refining,li2018qtcp,nascimento2019drl}.

More specifically, the literature is much less abundant in the scope of packetized fronthaul networks. In this context, 
\ac{RL} approaches have the additional challenge of dealing with very high-speed (microsecond) control loops, which impose stringent
 requirements on the inference time of Neural Network \ac{NN} models. If the inference time is slower than the minimum 
 \acp{RTT}, \ac{RL} agents cannot be very responsive, which could deteriorate their performance~\cite{abbasloo2020sigcomm}. Another 
 usual concern regarding \ac{RL} models is their black-box nature. Their low interpretability hides from the practitioner the 
 decision-making process, posing questions on how general the models are, and if they can be broadly trusted.

Overall, those are valid questions, but although RL models usually perform very well in scenarios similar to those observed during the 
training phase (in training distribution), the open question remains on how to obtain the same guarantee in scenarios unseen during the 
training phase~\cite{kirk2021generalization}. 

% One of the main challenges regarding the generalization capabilities of \ac{RL} models, is that of performing well even in 
% scenarios that were not seen during the trainig phase. is a vast area of research 

% This paper proposes a methodology to circumvent all the aforementioned
% challenges. First, to deal with performance issues of general \ac{CC}, we train
% a \ac{RL} baseline policy specialized in fronthaul-like scenarios. Second, to
% deal with inference-time and interpretability issues, we employ deep symbolic
% regression to extract closed-form mathematical expressions (symbolic policies)
% that approximate the \ac{RL} baseline behavior. The resulting symbolic policies
% are interpretable, and at the same time, easy to implement in any programming
% language, which completely overcomes any issues with inference time. The
% results show that such policies also follow very closely the overall
% performance of the \ac{RL} baseline, while matching its generalization
% capabilities.

This paper proposes a methodology to circumvent all the
challenges mentioned earlier. First, to deal with performance
issues of general \ac{CC}, we train a \ac{RL} baseline policy specialized
in fronthaul-like scenarios. Second, to deal with inference-
time and interpretability issues, we employ deep symbolic
regression to extract closed-form mathematical expressions
(symbolic policies) that approximate the \ac{RL} baseline behavior.
The resulting symbolic policies are interpretable and at the
same time, easy to implement in any programming language,
which completely overcomes any issues with inference time.
The results show that such policies also closely follow
the overall performance of the \ac{RL} baseline while matching
its generalization capabilities.

% This approach, makes inference-time
% negligible and exposes the inner workings of the congestion control policy as a function of the
% input variables, making it completely explainable. 

% The method consists of training a \ac{TD3} policy
% to act as a baseline and then via regression clone the behavior of such agent. A comprehensive set
% of experiments compares the performance of the original \ac{TD3} policy and its symbolic
% counterparts. 

% The paper is organized as follows. Section~\ref{sec:background} provides a background on \ac{RL},
% symbolic regression and congestion control. Section~\ref{sec:problemdesc} describes the training
% environment with its underlying learning problem, technical aspects of the network simulations, and
% the methodology for learning the symbolic policies. Finally, Section~\ref{sec:experiments} describes
% the experiments and results for different network configurations, and Section~\ref{sec:conclusions}
% presents our concluding remarks.

The paper is organized as follows. Section~\ref{sec:background} presents symbolic regression, RL, and congestion control 
background. Section~\ref{sec:problemdesc} describes the learning problem, training environment, network simulation technical 
aspects, and the methodology for learning the symbolic policies. Section~\ref{sec:experiments} describes the experiments and 
results for different network configurations, and Section~\ref{sec:conclusions} presents our concluding remarks.
%
% One simple split on the issue is between generalizing
% in \ac{IID} or \ac{OOD} scenarios. In the fomer case, the objective is that the agent is able
% to generalize across unseen input data that are sampled from the same training distribution, while
% in the later the focus is for the agent to genelize across samples from other distributions.%,
% Another important aspect, specially for ODD case, is if such
% generalization occurs in a Zero-shot Policy Transfer fashion, that is, no extra fine tunning
% is required on the model for it to perform well in the target environment.

%% file: sections/background.tex
\section{Background}\label{sec:background}

\subsection{Reinforcement Learning}

\ac{RL} comprises a set of techniques that enable an agent to learn how to
optimally interact with an environment via trial-and-error. The overall goal is
to derive policies that map the current observed state $x_t\in
\mathcal{X}\subseteq\mathbb{R}^n$ of a system and produce actions
$a_t\in\mathcal{A}\subseteq\mathbb{R}$ that maximize the expected cumulative
rewards $r: \mathcal{X}\mapsto \mathbb{R}$ received over
time~\cite{sutton1998introduction}.

%At$ every timestep $t=1,2,\dots$, the agent observes the current state of the system $s_t\in
%\mathcal{S}$ and by consulting its policy $\pi: \mathcal{S}\mapsto\mathcal{A}$ decides for an action
%$a\in \mathcal{A}$ that induces a state transition to $s_{t+1}$. As result of such transitions, a
%reward $r: \mathcal{S}\mapsto\mathbb{R}$ is produced which indicates if the action was beneficial or
%detrimental. An optimal policy $\pi*$ induces state transitions $\tau=s_1,\dots,s_{H}$ such
%$R(\tau)=\sum_{s\in\tau} r(s)$ is maximum. For alternative formulations
%see~\cite{sutton1998introduction}.

%There are different \ac{RL} techniques to deal with the many aspects that arise in practice. In some
%environments, the agents must choose from a finite set of actions, while in others, they face nearly
%infinite-sized action spaces. The same is true for the state spaces, ranging from small grid-worlds
%to continuous state spaces. In both cases, models for function approximation have become a
%requirement to deal with real-world environments, a role mainly fulfilled by Deep \acp{NN}, hence
%\ac{DRL}. 

%There are many categories of problems and solution methods for RL that depend on aspects such as the
%existence of single or multiple agents, the actions being discrete, or continuous, among others.
%This paper focuses on single-agent training environments with discrete action space and learning
%mechanisms based on Temporal Difference (TD) learning~\cite{sutton1998introduction}.

This paper focuses on \ac{RL} algorithms suitable to environments where the
action space is continuous. As a representative example, we describe the
\ac{DDPG}~\cite{lillicrap2019continuous}, which employs an actor-critic
architecture combining both Q-learning and policy gradient
techniques~\cite{sutton1998introduction}. 

The actor is a deterministic policy $\mu_{\theta}: \mathcal{X} \mapsto
\mathcal{A}$ that, given a state, outputs actions in a continuous-space, while
the critic is a parameterized Q-value function $Q_{\phi}:\mathcal{X}\times
\mathcal{A}\mapsto \mathbb{R}$ that assess the quality of such actions in terms
of expected future rewards. \ac{DDPG} follows an off-policy strategy which
relies on an experience buffer $\mathcal{D}=\{(x_t, a_t, x_{t+1},
r(x_t))_i\}_{i=0}^\ell$, containing action-induced state
transitions~\cite{mnih2015human}.

% Training of the critic, with parameters $\phi$, is based on \ac{TD} learning and the minimum squared
% error, similarly to \ac{DQN}. However, instead of computing the action that leads to maximum
% $Q_{\phi'}$, it uses the value for the action provided by the target policy
% $a_{t+1}:=\mu_{\theta'}(s_{t+1})$. 
%\begin{align}
    %\mathcal{L}_\mathcal{D}(\phi) = \mathop{E}_{e\sim \mathcal{D}} \left[ \left(Q_\phi(s_t,a) - (r_t +
    %\gamma Q_{\phi'}(s_{t+1}, a_{t+1}))\right)^2\right]
%\end{align}
%
% By fixing the parameters of the critic ($\phi$), the actor, with parameters, $\theta$, can be trained
% via the deterministic policy gradient algorithm, towards actions that maximize
% $Q_\phi$~\cite{silver2014dpg}.
%\begin{align}
    %\theta = \arg\max_\theta \mathop{E}_{s\sim \mathcal{D}} \left[Q_\phi(s_t, \mu_\theta(s_t))\right]
%\end{align}
% \ac{DDPG} updates its target value-functions and policies via a procedure
% known as Polyak averaging that aggregates all previous versions of the
% \acp{NN}~\cite{lillicrap2019continuous}.

To address \ac{DDPG} shortcomings regarding the overestimation of
Q-values~\cite{hasselt2016drl}, the \acf{TD3} algorithm was
proposed~\cite{fujimoto2018addressing}, introducing: (1) clipped double
Q-Learning, (2) delayed policy updates (and target networks), and (3) target
policy smoothing. \ac{TD3} is the baseline \ac{RL} agent used for the
experiments in this paper.

% Generalization
%% Jean: Moved this discussion to the introduction, in a simplified version
% Differently from most \ac{SL} approaches, where generalization is translated into model performance
% in a test data set, in \ac{RL} generalization can take the form of a more complex problem and has not been
% so extensively study~\cite{kirk2021generalization}.
% In RL, generalization can be seen as a class of problems where these classes
% can also take many formats. One simple split on the issue is between generalizing
% in \ac{IID} or \ac{OOD} scenarios. In the fomer case, the objective is that the agent is able
% to generalize across unseen input data that are sampled from the same training distribution, while
% in the later the focus is for the agent to genelize across samples from other distributions.%,
% Another important aspect, specially for ODD case, is if such
% generalization occurs in a Zero-shot Policy Transfer fashion, that is, no extra fine tunning
% is required on the model for it to perform well in the target environment.
%
%\begin{figure}[!htb]
%  \centering
%  \includegraphics[width=0.4\textwidth]{figs/gen1.png}
%  \caption{Generalization perspectives: (a) \ac{IID} (b) \ac{OOD}}
%  \label{fig:generalization}
%\end{figure}

\subsection{Deep symbolic regression}

Given a dataset $\mathcal{D}= \{X_i, y_i\}$, where every $X_i\in \mathbb{R}^n$
and labels $y_i\in\mathbb{R}$, symbolic regression is a supervised learning
procedure that aims at identifying a function $f: \mathbb{R}^n\mapsto
\mathbb{R}$ that minimizes the residual $||f(X_i) - y_i||^2$ and whose form is
a short mathematical expression defined from a set of predefined tokens. If the
dataset $\mathcal{D}$ represents the states observed and the actions taken by
an \ac{RL} policy, then function $f$ is a white-box approximation of such
policy. This type of procedures are usually referred to as behavioral cloning
of the expert policy~\cite{abdoulaye2021behavior}. However, depending on the
properties of the dataset and the accuracy of the regression process,
the expert RL policy and the resulting $f$ may show very different
capabilities~\cite{udrescu2020aifeynman}. 

In this paper, we employ the deep symbolic regression method proposed by
Petersen et
al~\cite{petersen2021deep}\footnote{\url{https://github.com/brendenpetersen/deep-symbolic-optimization}}.
The method employs a neural network controller to represent a distribution over mathematical expressions defined as a sequence of tokens in the pre-order of the corresponding expression tree. As an example, the sequence $\tau=[+, \sin, x_1, \exp, x_2]$ corresponds to the expression $f(x)=\sin(x_1) + \exp(x_2)$. 

Expressions are sampled autoregressively from the probability distribution learned by the controller, and they are evaluated based on how well they match the dataset $R(\tau, \{X_i, y_i\})$. The controller is then trained to maximize a \ac{RL} objective based on $\mathbb{E}_\tau=[R(\tau, \{X_i, y_i\})]$. As the training progresses, the controller is able to generate expressions that better match the dataset~\cite{landajuela2022a}.

%% file: sections/methodology.tex
\section{Methodology}%
\label{sec:problemdesc}

% This section defines the methods required to evaluate if policy distillation is a effective way of
% dealing with impeditive inference time of \acp{NN} ($ms$) in fast control loops ($\mu s$), such as
% those found in fronthaul networks. First, we describe the simulation environments where the \ac{TD3} agent is
% trained and the underlying learning problem, next we define the policy distillation process
% that converts the trained \ac{TD3} policy into decision trees. 

%Training of the TD3 agent produces a policy $\pi_\theta: o\mapsto a$ that outputs an action for a given observation as input. The time required for such a policy to output an action depends on the hardwared employed for inference, but for practical reasons it is natural to assume that no 

\subsection{Network simulations}

\begin{figure}[htb]
    \centering
    \includegraphics[width=.75\columnwidth]{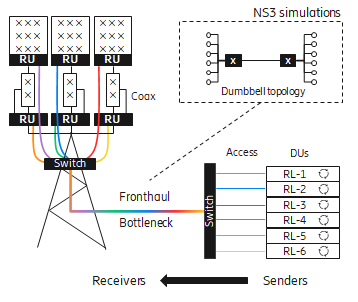}
\caption{\hl{Fronthaul network and its NS3 counterpart.}}%
    \label{fig:topology}
\end{figure}

The network environment is a fronthaul simulation developed on \ac{NS-3}, with
additional support to OpenGym interfaces for RL implemented using
ns3-ai~\cite{hao2020ns3ai}. The fronthaul scenario implements a UDP-based
constant bit-rate communication between pairs of senders (DUs) and their
receivers (RUs), which share a bottleneck link in the dumbbell topology of
Figure~\ref{fig:topology}. Senders and receivers are connected to switches
through individual access links; and these switches exchange packets via the
shared communication link. To isolate the dynamics we want to explore, we
assume the access links have negligible packet losses and sufficient capacity,
so that these impairments are only present in the shared link. To perform the
congestion control on top of the UDP-based communication, agents are given
control of the intersend time between sent packets, implementing a rate-based
congestion control approach.

%\begin{figure}[!h]
%     \centering
%     \includegraphics[width=0.45\textwidth]{figs/ns3gym.png}
%     \caption{ns-3 environment implementation.}
%     \label{fig:ns3env}
% \end{figure}

Typically, congestion control algorithms are event-based, in that they act upon
the occurrence of a trigger event, for instance the reception of an ACK or
detection of a lost packet from a timeout. The RL agent, on the other hand, was
implemented as a time-based algorithm. This allows the agent to observe the
impact an action for a while before a new action is taken. Therefore, the
observation time window is also a simulation parameter that needs to be defined
and influences the overall performance of the
agent~\cite{zhang2020reinforcement,abbasloo2020sigcomm}. 

%From an RL perspective, the proposed analysis required a scalable robust
%execution environment able to cope with the large number of simulations needed
%to train the agent. As such, an execution environment was setup on an OpenStack
%cluster, employing the RAY framework~\cite{moritz2018} as the main tool for
%managing the distributed computation and experiment life-cycles. As for the
%TD3 agent, we used the implementation available on RLlib~\cite{liang2017rllib}.

% Because the network is simplified in a dumbbell topology, the ratio between the
% total bandwidth required by all nodes and bottleneck size is a key parameter,
% and different bottleneck configurations were evaluated.
% Another important parameter is the duration of each simulation ---
% in our case, each a simulation episode runs for 3~s during training.
%
\subsection{Learning Problem}

%Todo:
%\begin{itemize}
%    \item How many learning agents? One RL.
%    \item Reward signal: global vs local? local
%    \item How often to observe? Every timestep $t$
%    \item How tight is the bottleneck?
%    \item How these possibilities interact?
%    \item Observation and action spaces
%    \end{itemize}

The network simulations define a multi-agent scenario where senders must
cooperate in order to maximize link utilization and fairness while minimizing
the RTTs. Here, we assume a decentralized learning problem where the agents
cannot communicate, and only observe their local performance metrics, utilities
and rewards. Finally, we also assume that all agents follow the same learned
policy.

The observation space is defined by four dimensions
$\mathcal{X}\subset\mathbb{R}^4$. At any timestep $t$, we define $x_1^t$ as the
intersend time, $x_2^t$ as the average RTT (observed during the current time
window), $x_3$ as the RTT ratio ($x_2$ over the minimum observed RTT), and
$x_4^t$ as the packet loss ratio (packet losses over the number of packets
sent). 

The action space is unidimensional $\mathcal{A}=[0.8,1.5]$, and the actions $a_t\in\mathcal{A}$ are employed to 
to update the intersend time as follows $x_1^{t+1} := x_1^{t}/a_t$.

% Finally, the reward function was defined based on three measurements: the ratio between current
% round trip time over the minimum observed RTT, the ratio between number of losses over all packets
% sent during the timestep and the ratio of acknowledged packets over the maximum number of packets
% that could be sent with the current timestep when using the full bandwidth.

The reward is a linear function of the number of acknowledged packets (acks$_t$), the average round trip time ($RTT_t$), and packet losses (losses$_t$). We assume all measurements are first normalized to the same scale by a function $\eta$. Since the reward function is not employed during the evaluation, the normalization function can rely on information that is not accessible after training, e.g., the bounds for number of acknowledged packets ($A$), RTTs ($R$), and packet losses ($L$).
% \begin{align}
%   r_{t} =  \cdot \left( \frac{\text{acks}}{\text{acks}_{\text{max}}}
%   - \frac{\widehat{\text{RTT}}}{\text{RTT}_{\text{\text{min}}}}
%   - \frac{\text{losses}}{\text{losses}_{\text{max}}} \right)
% \end{align}
\begin{align}
    r_{t} =  \eta(\text{acks}_t,A) - \eta(RTT_t,R) - \eta(\text{losses}_t,L)
    \label{eq:rewards}
\end{align}

% {\color{red} Should add some words about the consequences of this sum,
%   regardless of normalization. Maybe something like ``to balance the terms, we
%   must calculate, for example, what an x\% increase to RTT ratio does to the
%   bitrate as calculated by the receiver''.}

Equation~\eqref{eq:rewards} aims at inducing the trained policies to
increase the transmission rate until the observed RTTs increase or packet losses occur.
Overall, an optimal policy would allow a sender to discover the highest fair transmission rate 
that maintains the RTTs close to the minimum. 
% \begin{figure}[!h]
%     \centering
%     \includegraphics[width=0.9\columnwidth]{figs/reward.png}
%     \caption{The agents are rewarded for increasing their throughput but also penalized if the RTTs
%     or packet losses increase.}%
%     \label{fig:rewards}
% \end{figure}
%
The above formulation models congestion control as a decentralized
decision-making problem. It is a harsh environment for learning, due
to partial observability and the inherent non-stationarity of the network
state, but it makes the scenario close to real-world deployments.

% While desirable, full observability would require the
% introduction of additional control messages to propagate the information through
% the network, further increasing the competition for resources.

%\subsection{Training setup}

%This section should describe aspects of training agents for multi-agent environments by using
%parameter sharing. APEX TD3 from RLlib, to speed up training. The output of this section is a TD3
%agent, the next section transforms it in DTs. 

\subsection{Deep symbolic regression}
\label{sec:dsr}

Given an expert \ac{TD3} policy $\mu_\theta: \mathcal{X}\mapsto\mathcal{A}$,
the first step of learning a closed-form congestion control policy is to
collect a dataset of experiences from $\mu_\theta$ in the target environment.
Such experiences are then stored as state-action tuples in a dataset
$\mathcal{D}=\{(x_i, a_i)\}_{i=1}^N$, where $x_i\in \mathcal{X}$ are
observations from the current state and $a_i\in\mathcal{A}$ are the respective
actions proposed by the expert policy, $a_i:=\mu_\theta(x_i)$. 

Such datasets must contain a representative set of experiences. In congestion
control tasks, that means samples representing scenarios where the transmission
rate should be increased, stabilized, and decreased. To collect samples
of such classes, we employ a simple $\epsilon$-greedy exploration strategy,
with $\epsilon=0.5$, in which \ac{TD3} actions are chosen half of the time, and
random actions chosen otherwise. This simple exploration strategy eventually
leads the simulations to states of high RTT values and packet losses.

\begin{figure}[htb]
    \centering
    \includegraphics[width=.49\textwidth]{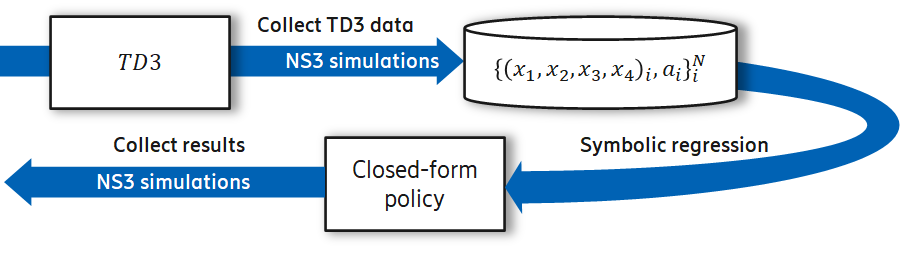}
    \caption{The symbolic regression setup consists of two steps (1) data collection and
    (2) deep symbolic regression.}%
    \label{fig:regression}
\end{figure}

From a data set $\mathcal{D}$, and a set of predefined tokens $\tau$, and a maximum
length $\ell$, a symbolic regression task can be specified whose goal is to
learn a closed-form expression capable of approximating the expert decisions (see
Figure~\ref{fig:regression}).

The closed-form expressions (or symbolic policies) that result from this
process, will achieve different performances depending on their ability to imitate
the expert actions in both in-training and out-training distributions.

%% file: sections/results.tex
\section{Experiments and results}%
\label{sec:experiments}

This section compares the learned closed-form congestion control policies to an
RL baseline. We assess performance regarding link utilization, packet delay,
packet losses, and fairness. We assess generalization capacity by evaluating
network scenarios not included in the training datasets.

\subsection{Reinforcement learning baseline}

The training environment for the \ac{TD3} baseline consists of a set of different
network scenarios, defined by their bottleneck capacity and the number of
sender-receiver pairs. We employ domain randomization over the bottleneck capacity and the number of
sender-receiver pairs~\cite{tobin2017domain}, according to Table~\ref{tab:env}. 

\begin{table}[h]
    \centering
    \caption{Network simulation training environment.}
    \begin{tabular}{l|l}
        Parameter & Domain \\\hline
        Bottleneck capacity & $1Mbps \leq C \leq 2Gbps$\\
        Number of senders & $p \in\{1,2\}$ \\
        Access links & $20Gbps$ (overdimensioned) \\
        Switch queue size & $100$ packets \\
        Simulation duration & $3$s \\
        Timestep & $1ms$\\
    \end{tabular}%
    \label{tab:env}
\end{table}

The hyperparameters for the \ac{TD3} algorithm followed those employed by
\cite{martins2021policy}. The actor and critic neural networks architectures as
$512 \times 16 \times 512$, actor learning-rate \num{1e-5}, critic
learning-rate \num{5e-5}, multi-step learning with $n=5$, and $\gamma = 0.999$
(see the reference for more details on this training setup).

\subsection{Closed-form congestion control policies}

The symbolic regression method described in \ref{sec:dsr} produces closed-form
expressions that approximate the output label described in the datasets. To
stress the generalization capabilities of such policies, we utilize a minimal
set of network scenarios for data collection, with a single bottleneck capacity
of $C=500Mbps$.

Following this setup, we collected three different datasets using different
numbers of sender-receiver pairs $p$, here denoted as $\mathcal{D}_{p=1}$,
$\mathcal{D}_{p=2}$, and $\mathcal{D}_{p=1,
2}=\mathcal{D}_{p=1}\cup\mathcal{D}_{p=2}$. Each dataset contains experiences
from 5 seconds NS3 simulations.

Each dataset is input to a deep symbolic regression method, which aims to
produce closed-form expressions of maximum length $\ell=32$ tokens. The set of
available tokens is limited to $\tau=\{+,-,\times,\div,\cos\}$. After a fixed
amount of time, all runs were stopped, and the best-ranked expression was chosen
for the evaluation experiments, which we describe below.

\begin{align}
    \pi_{\mathcal{D}_{p=1,2}} &= \cos{\left(x_{2} + \frac{2 x_{2}}{\frac{x_{1}}{x_{3}^{3} + \frac{- x_{1} + x_{2}}{x_{1}}} + x_{2}} \right)}\tag{SP1}\\
    \pi_{\mathcal{D}_{p=1}} &= - \frac{x_{3}}{x_{3} + \frac{\frac{x_{1} x_{3}}{x_{2}^{2}} + x_{3}}{x_{2} x_{3}}} + \frac{\cos{\left(\frac{x_{2}}{x_{1}} \right)}}{x_{3}^{2}}\tag{SP2}\\
    \pi_{\mathcal{D}_{p=2}} &= \cos{\left(\frac{x_{2} x_{3} \left(x_{2} x_{3} + x_{2} + 2 x_{3} + x_{4}\right)}{x_{1} + x_{2}^{2} x_{3} x_{4} + x_{2} x_{3}^{2}} \right)}\tag{SP3}
\end{align}

For reference, $x_1, x_2, x_3, x_4$ stands for inter-send time, RTT, RTT ratio,
and packet loss ratio. It is worth noticing that SP1 and SP2 have ignored the
information about packet losses, i.e., $x_4$. This is an indication that to
approximate the actions in the dataset, such variable was not essential.

Another intriguing characteristic is the presence of tokens such as $\cos$,
whose relevance was identified in preliminary experiments. It introduces a
non-linearity that showed benefits in most of the evaluated regression settings.

% Regarding the maximum length of the expressions, we found $\ell=32$ as a good
% trade-off betwen complexity and performance, but a smaller $\ell$ to produce
% even simpler expressions.
%

\subsection{Performance Evaluation}

Here we compare the performances of \ac{TD3} and the symbolic policies SP1,
SP2, and SP3. We design a two-phase set of experiments, where Phase I identifies
the most promising symbolic policy in short simulations, and Phase II performs more expensive simulations 
against the \ac{TD3} baseline (Table~\ref{tab:eval}).

\begin{table}[h] 
    \centering
    \caption{Evaluation settings.}
    \begin{tabular}{l|l|l}
        Parameter& Phase I (1s)& Phase II (20s)\\\hline
        $C\in $ (Mbps) & $250, 500, 1000$ & $250, 500, 1000$\\
        $p\in$ & $\{1,5,10,15,20\}$ & $\{1,5,10,\dots,35,40\}$\\
    \end{tabular}%
    \label{tab:eval}
\end{table}

The different bottleneck capacities induce network scenarios with different RTT
ranges, while the number of senders impacts the network dynamics. An optimal
agent would maintain high link utilization (close to 100\% of the bottleneck
capacity), average RTTs close to the minimum, and zero packet losses while
guaranteeing fair bandwidth shares for each flow. It is also desirable that the
trained policies generalize well to a superset of the scenarios seen during
training. Therefore, the results here contain simulations with as many as 40
sender-receiver pairs, much more than those observed during training.

To get overall performance picture, we normalized and aggregated all the
measurements. Additionally, since there were never packet losses, we only include 
plots for RTT, Jain fairness index and link utilization.

\begin{figure}[h]
    \centering
    \subfloat[Phase I.]
    {
        \includegraphics[width=.95\columnwidth]{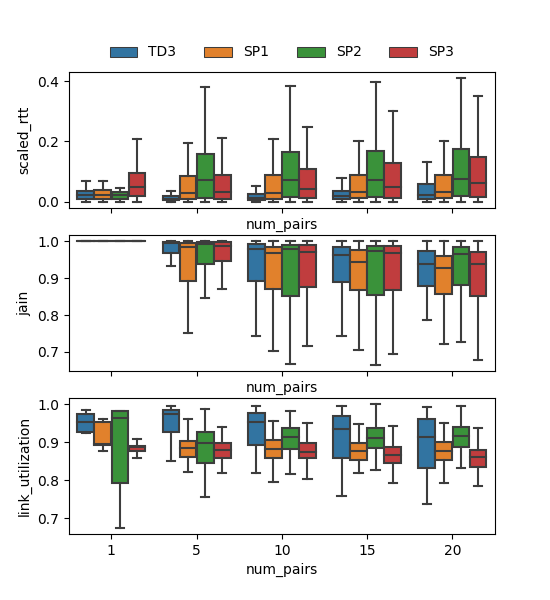}
        \label{fig:perfI}
    }\\[-0.09cm]
    \subfloat[Phase II.]
    {
        \includegraphics[width=.95\columnwidth]{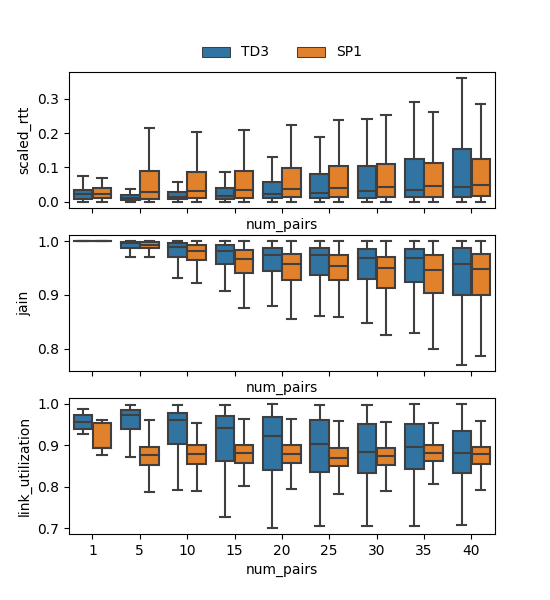}
        \label{fig:perfII}
    }
    \caption{Performance results of Phases I and II.}
\end{figure}

Figure \ref{fig:perfI} summarizes the results of Phase I. The results
illustrate performance in two scenarios, those close to the training
distribution and those far from it. In the first scenario, we highlight the
case of $p=1$, that belongs to the training distribution of SP1 and
SP2 but not for SP3, where SP3 reached the worst results overall. Conversely,
in scenarios with $p\geq 5$ that are out of the training distribution for all
symbolic policies, we observe a more stable behavior. 

Overall, SP1 and SP3 performed better than SP2 in terms of RTT, which relates
to the fact that they are also more conservative regarding link utilization.
We choose SP1 to go for Phase II. 

Figure~\ref{fig:perfII} summarize the Phase II results of SP1 against TD3.
Regarding RTT and link utilization, we observe an interesting pattern in which
SP1 performance deteriorates slower than TD3 as $p$ increases. This is an
impressive result that provides evidence for high generalization capacity of
the symbolic policies. Regarding the Jain fairness index, all the policies variances
increased very similarly with $p$. Regarding link utilization, we again see a
slower deterioration of SP1 performance when compared to TD3.

%generalization
%% Jean: I merged this in the beginning of the section
% As part of the performance evaluation, we also analyzed the policies capabilities
% of generalization over the sender-receiver pairs dimensions. During training, agents
% only observes experiqences from 1 or 2 sender-receiver pairs, while during evaluation
% this number is increased up to 40 pairs. 
% The agent's observation space is
% a four dimensional vector composed of average rtt, ratio of packets loss over sent,
% current intersend time and ratio between average rtt and minimum observed RTT. Some values,
% such as the minimum observed RTT, are extremely dependent on the network
% conditions. Increasing the number of sender-receiver pairs significantly the agent
% is exposed to unseen scenarios and, as illustrated in Figure~\ref{fig:perf}, it
% was able to generalize well to these new conditions, indicating the agents ability
% to generalize in OOD scenarios in a zero-shot policy transfer fashion.

\subsection{Analysis and visualization}

We turn now to a brief investigation on the symbolic policy's ability to cope
with different scenarios and network conditions. Due to space constraints, we
focus only on SP1 and a few examples that illustrate our main points.
Furthermore, we work around the challenges introduced by a four-dimensional
observation space with convenient slices, as will be explained in the next
paragraphs.

\begin{figure}[t]
 \centering
 \includegraphics[width=.9\columnwidth]{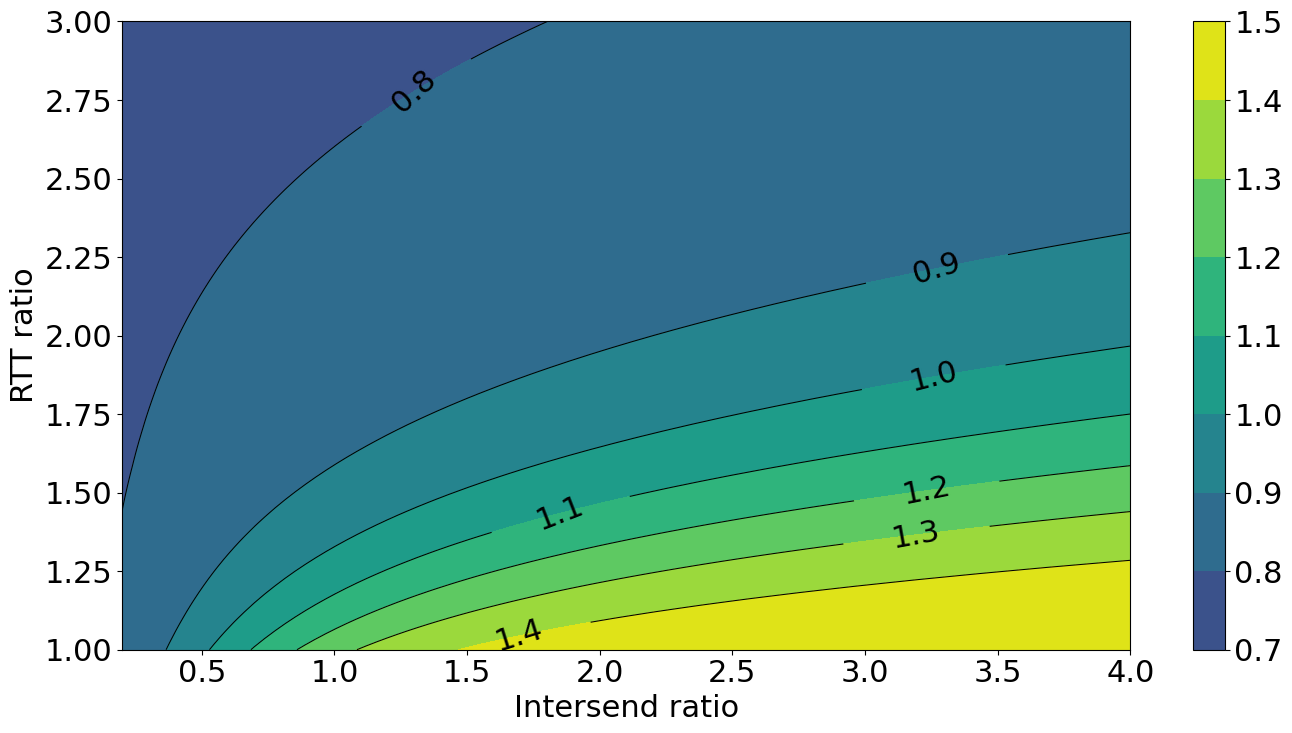}
    \caption{Contours of simplified SP1 for $C=1000Mbps$.}%
 \label{fig:heatmap}
\end{figure}

In line with previous work on visualizing such models~\cite{martins2021policy},
Figure~\ref{fig:heatmap} presents contours of the output of SP1, remapped to
action space $\mathcal{A}$. The axes are ``Intersend ratio'' and ``RTT ratio''.
Indeed, it is possible to rewrite SP1 in terms of a minimum RTT $c$ (primarily
defined by link speeds in the topology):
\begin{align}
  \label{eq:rewrite_sp1}
  f(i_{\text{ratio}}, \text{rtt}_{\text{ratio}}) &= \
    n(\pi_{\mathcal{D}_{p=1,2}}(i_{\text{ratio}}\cdot c, \text{rtt}_{\text{ratio}}\cdot c, \text{rtt}_{\text{ratio}})),
\end{align}
where $x_{1}=i_{\text{ratio}}\cdot c$ is the observed packet intersend time and
$x_{2}=\text{rtt}_{\text{ratio}}\cdot c$ is the observed RTT, in terms of
$x_{3}=\text{rtt}_{\text{ratio}}$, and $n(\cdot)$ simply converts the [-1,1]
range to $\mathcal{A}$.

With Eq.~(\ref{eq:rewrite_sp1}), we can interpret the horizontal axis of
Figure~\ref{fig:heatmap} as the agent's \emph{offered load} -- normalized to the
minimum RTT of the network --, where a value of 0.5 would signify sending 2
packets of data per round-trip time, and a value of 4.0 would denote a much less
``aggressive'' sender that issues a single packet of data every four
round-trip times. This can be easily translated to a bit-rate, e.g. in Mbps, if the
data packets are of the same size. In the vertical axis, one can explore
several \emph{network conditions} via the observed RTT ratio, with a value of
1.0 signifying empty queues in the forward and reverse paths 
%(i.e., one-way delays are only due to serialization times), 
and a value of 2.0 denoting double the minimum RTT.

We draw attention to the 1.0 contour line in Figure~\ref{fig:heatmap}, which
indicates that the offered load should not be changed. Above that, the darker
regions on higher RTT ratio values have outputs that \emph{increase} the
intersend time, lowering the offered load; whereas the brighter yellow regions
on higher intersend ratios indicate that agents with presumably small offered
loads can increase them more dramatically than agents whose packets leave more
frequently. We note that the heatmap for a scenario with a 100-Mbps bottleneck
is very similar, indicating that this ``fairness'' behavior is also present on
more capacity-limited scenarios, though we elide that figure due to space
constraints.

%As mentioned previously, 
It is not immediately clear what benefits the
introduction of a cosine function brings to the policy's performance. With this
in mind, we investigate SP1's output with respect to several intersend ratio
values, with slices across different RTT ratios and bottleneck link speeds (as
defined by a minimum RTT).

\begin{figure}[t]
 \centering
 \includegraphics[width=.9\columnwidth]{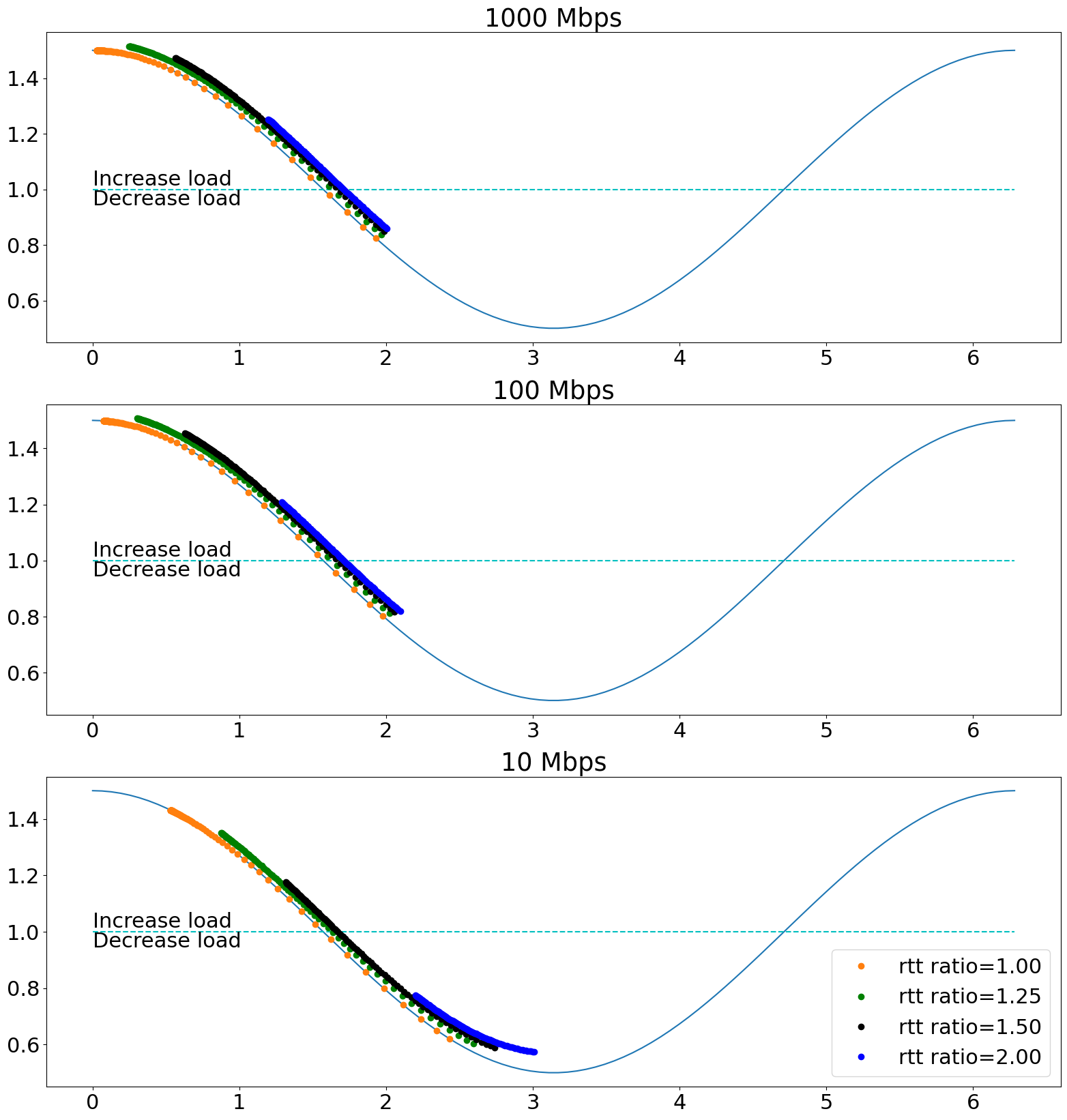}
    \caption{Span of SP1 policy values over the cosine domain.}%
 \label{fig:cosines}
\end{figure}

This is depicted in Figure~\ref{fig:cosines}, where the simplified SP1 is
plotted against the [0,$2\pi$] domain of the cosine function. In this figure,
the intersend ratio varies in the [0.2,10.0] range, and we artificially separate
the dots from each slice vertically, for ease of understanding. One can see
that, for the $\text{rtt}_{\text{ratio}}=1.0$ slice, i.e.\ no queueing in the
network, most of the actions are in the ``Increase load'' region though,
understandably, this behavior becomes less prominent as the bottleneck link
capacity gets smaller. In general, this figure illustrates that the policy
adapts its output for different scenarios and conditions by evaluating different
regions of its underlying non-linear functions.

\begin{figure}[hbt]
 \centering
 \includegraphics[width=.8\columnwidth]{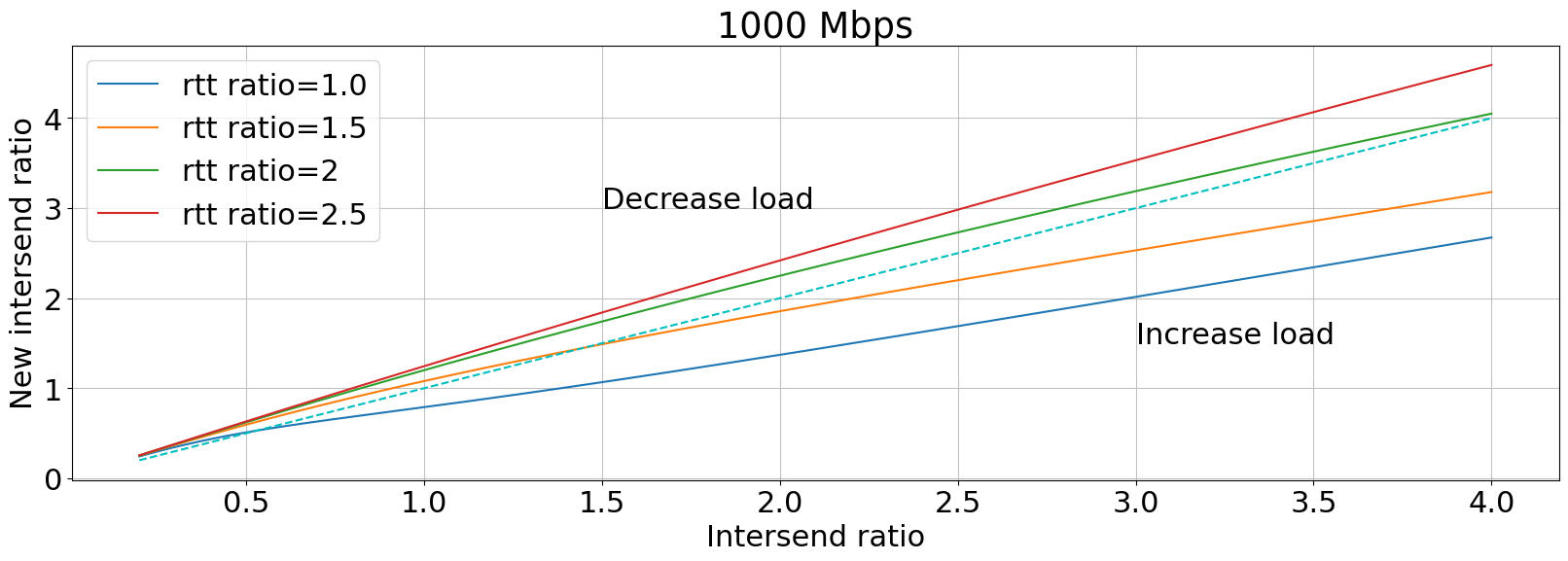}
 \includegraphics[width=.8\columnwidth]{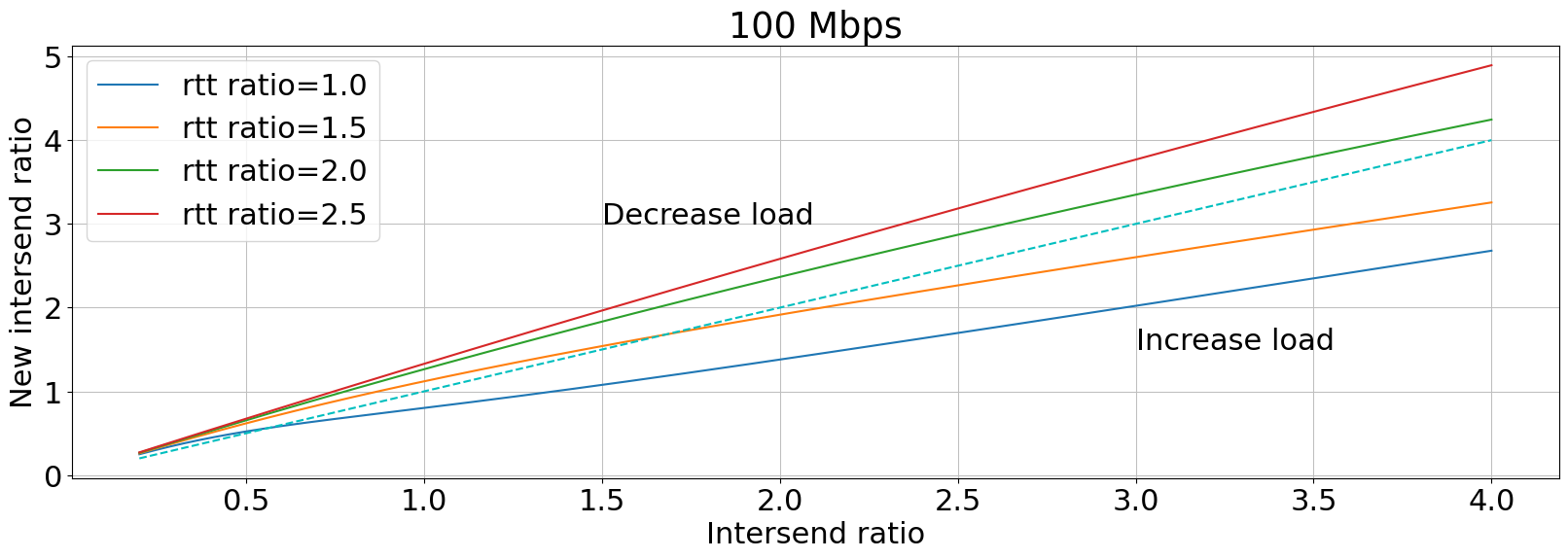}
 \includegraphics[width=.8\columnwidth]{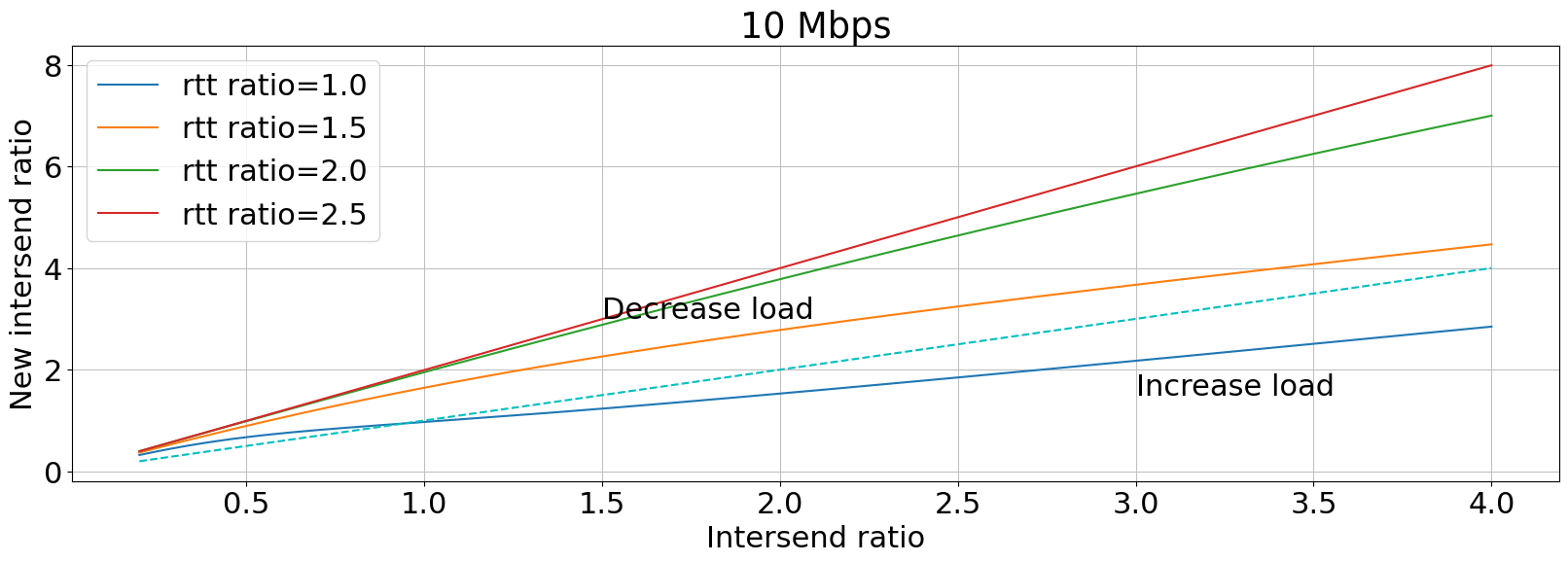}
    \caption{New intersend ratio for several bottleneck scenarios.}%
 \label{fig:ir_newir}
\end{figure}

% NOTE (igor): i'm not entirely happy with this part, we can probably remove it
As an extension of Figure~\ref{fig:cosines}, we plot below the new intersend
ratio values an agent would employ after evaluating the SP1 policy -- unlike the
previous figures, the no-change region in Figure~\ref{fig:ir_newir} is the
dashed $x=y$ line. Once more, we draw attention to the similarity in the outputs
for scenarios with very different capacities, and the general trend of avoiding
excessive load in extremely limited scenarios.

%% file: sections/conclusions.tex
\section{Conclusions}%
\label{sec:conclusions}

Recent results have shown that specialized \acf{RL} congestion control policies 
are effective alternatives in cases where traditional algorithms do not perform well.
However, the deployment of \ac{NN} models comes with challenges regarding performance, 
inference time, and generalization guarantees. In the context of ultra-low-latency 
packetized fronthaul networks, for example, a low inference time is a hard requirement 
that might restrict the applicability of certain policies in the real world.

This paper proposed and evaluated the use of deep symbolic regression for overcoming 
inference time challenges while enabling model interpretability and maintaining reasonable
generalization capabilities. We trained a fronthaul-specific congestion control policy via \ac{RL}, 
and then employed deep symbolic regression on small state-action datasets collected
from the \ac{RL} baseline experiences. This process produced closed-form symbolic policies whose 
output approximates the actions output by \ac{TD3}.

The results confirmed that the resulting closed-form policies could maintain a performance
 very similar to that of \ac{RL} baseline both in- and out-distribution of training data, which
means significant generalization capabilities. Additionally they resolve any eventual issue 
with inference time since they can be directly implemented in any programming language.

%% file: main.bbl
\begin{thebibliography}{10}
\providecommand{\url}[1]{#1}
\csname url@samestyle\endcsname
\providecommand{\newblock}{\relax}
\providecommand{\bibinfo}[2]{#2}
\providecommand{\BIBentrySTDinterwordspacing}{\spaceskip=0pt\relax}
\providecommand{\BIBentryALTinterwordstretchfactor}{4}
\providecommand{\BIBentryALTinterwordspacing}{\spaceskip=\fontdimen2\font plus
\BIBentryALTinterwordstretchfactor\fontdimen3\font minus
  \fontdimen4\font\relax}
\providecommand{\BIBforeignlanguage}[2]{{%
\expandafter\ifx\csname l@#1\endcsname\relax
\typeout{** WARNING: IEEEtran.bst: No hyphenation pattern has been}%
\typeout{** loaded for the language `#1'. Using the pattern for}%
\typeout{** the default language instead.}%
\else
\language=\csname l@#1\endcsname
\fi
#2}}
\providecommand{\BIBdecl}{\relax}
\BIBdecl

\bibitem{jaber20165g}
M.~Jaber, M.~A. Imran, R.~Tafazolli, and A.~Tukmanov, ``5g backhaul challenges
  and emerging research directions: A survey,'' \emph{IEEE Access}, vol.~4, pp.
  1743--1766, 2016.

\bibitem{holma2012lte}
H.~Holma and A.~Toskala, \emph{LTE advanced: 3GPP Solution for
  IMT-Advanced}.\hskip 1em plus 0.5em minus 0.4em\relax John Wiley \& Sons,
  2012.

\bibitem{chitimalla20175g}
D.~Chitimalla, K.~Kondepu, L.~Valcarenghi, M.~Tornatore, and B.~Mukherjee, ``5g
  fronthaul--latency and jitter studies of cpri over ethernet,'' \emph{Journal
  of Optical Communications and Networking}, vol.~9, no.~2, pp. 172--182, 2017.

\bibitem{jay2019deep}
N.~Jay, N.~Rotman, B.~Godfrey, M.~Schapira, and A.~Tamar, ``A deep
  reinforcement learning perspective on internet congestion control,'' in
  \emph{International Conference on Machine Learning}.\hskip 1em plus 0.5em
  minus 0.4em\relax PMLR, 2019, pp. 3050--3059.

\bibitem{zhang2020machine}
T.~Zhang and S.~Mao, ``Machine learning for end-to-end congestion control,''
  \emph{IEEE Communications Magazine}, vol.~58, no.~6, pp. 52--57, 2020.

\bibitem{congcontrol2021renaissance}
W.~Wei, H.~Gu, and B.~Li, ``Congestion control: A renaissance with machine
  learning,'' \emph{IEEE Network}, pp. 1--8, 2021.

\bibitem{xiao2019tcpdrinc}
K.~{Xiao}, S.~{Mao}, and J.~K. {Tugnait}, ``Tcp-drinc: Smart congestion control
  based on deep reinforcement learning,'' \emph{IEEE Access}, vol.~7, pp.
  11\,892--11\,904, 2019.

\bibitem{liu2019tyrus}
\BIBentryALTinterwordspacing
L.~Liu and H.~Xu, ``Tyrus: Phy-assisted neural adaptive congestion control for
  cellular networks,'' in \emph{Proceedings of the ACM SIGCOMM 2019 Conference
  Posters and Demos}, ser. SIGCOMM Posters and Demos '19.\hskip 1em plus 0.5em
  minus 0.4em\relax New York, NY, USA: Association for Computing Machinery,
  2019, p. 45–47. [Online]. Available:
  \url{https://doi.org/10.1145/3342280.3342302}
\BIBentrySTDinterwordspacing

\bibitem{abbasloo2020sigcomm}
\BIBentryALTinterwordspacing
S.~Abbasloo, C.-Y. Yen, and H.~J. Chao, ``Classic meets modern: A pragmatic
  learning-based congestion control for the internet,'' in \emph{Proceedings of
  the Annual Conference of the ACM Special Interest Group on Data Communication
  on the Applications, Technologies, Architectures, and Protocols for Computer
  Communication}, ser. SIGCOMM '20.\hskip 1em plus 0.5em minus 0.4em\relax New
  York, NY, USA: Association for Computing Machinery, 2020, p. 632–647.
  [Online]. Available: \url{https://doi.org/10.1145/3387514.3405892}
\BIBentrySTDinterwordspacing

\bibitem{eagle2020refining}
S.~Emara, B.~Li, and Y.~Chen, ``Eagle: Refining congestion control by learning
  from the experts,'' in \emph{IEEE INFOCOM 2020 - IEEE Conference on Computer
  Communications}, 2020, pp. 676--685.

\bibitem{li2018qtcp}
W.~{Li}, F.~{Zhou}, K.~R. {Chowdhury}, and W.~{Meleis}, ``Qtcp: Adaptive
  congestion control with reinforcement learning,'' \emph{IEEE Transactions on
  Network Science and Engineering}, vol.~6, no.~3, pp. 445--458, 2019.

\bibitem{nascimento2019drl}
I.~{Nascimento}, R.~{Souza}, S.~{Lins}, A.~{Silva}, and A.~{Klautau}, ``Deep
  reinforcement learning applied to congestion control in fronthaul networks,''
  in \emph{2019 IEEE Latin-American Conference on Communications (LATINCOM)},
  2019, pp. 1--6.

\bibitem{kirk2021generalization}
\BIBentryALTinterwordspacing
R.~Kirk, A.~Zhang, E.~Grefenstette, and T.~Rockt{\"{a}}schel, ``A survey of
  generalisation in deep reinforcement learning,'' \emph{CoRR}, vol.
  abs/2111.09794, 2021. [Online]. Available:
  \url{https://arxiv.org/abs/2111.09794}
\BIBentrySTDinterwordspacing

\bibitem{sutton1998introduction}
R.~S. Sutton and A.~G. Barto, \emph{Introduction to reinforcement
  learning}.\hskip 1em plus 0.5em minus 0.4em\relax MIT press Cambridge, 1998,
  vol. 135.

\bibitem{lillicrap2019continuous}
\BIBentryALTinterwordspacing
T.~P. Lillicrap, J.~J. Hunt, A.~Pritzel, N.~Heess, T.~Erez, Y.~Tassa,
  D.~Silver, and D.~Wierstra, ``Continuous control with deep reinforcement
  learning,'' in \emph{4th International Conference on Learning
  Representations, {ICLR} 2016, San Juan, Puerto Rico, May 2-4, 2016,
  Conference Track Proceedings}, Y.~Bengio and Y.~LeCun, Eds., 2016. [Online].
  Available: \url{http://arxiv.org/abs/1509.02971}
\BIBentrySTDinterwordspacing

\bibitem{mnih2015human}
V.~Mnih, K.~Kavukcuoglu, D.~Silver, A.~A. Rusu, J.~Veness, M.~G. Bellemare,
  A.~Graves, M.~Riedmiller, A.~K. Fidjeland, G.~Ostrovski \emph{et~al.},
  ``Human-level control through deep reinforcement learning,'' \emph{Nature},
  vol. 518, no. 7540, p. 529, 2015.

\bibitem{hasselt2016drl}
H.~v. Hasselt, A.~Guez, and D.~Silver, ``Deep reinforcement learning with
  double q-learning,'' in \emph{Proceedings of the Thirtieth AAAI Conference on
  Artificial Intelligence}, ser. AAAI'16.\hskip 1em plus 0.5em minus
  0.4em\relax AAAI Press, 2016, pp. 2094--2100.

\bibitem{fujimoto2018addressing}
\BIBentryALTinterwordspacing
S.~Fujimoto, H.~van Hoof, and D.~Meger, ``Addressing function approximation
  error in actor-critic methods,'' in \emph{Proceedings of the 35th
  International Conference on Machine Learning}, ser. Proceedings of Machine
  Learning Research, J.~Dy and A.~Krause, Eds., vol.~80.\hskip 1em plus 0.5em
  minus 0.4em\relax PMLR, 10--15 Jul 2018, pp. 1587--1596. [Online]. Available:
  \url{https://proceedings.mlr.press/v80/fujimoto18a.html}
\BIBentrySTDinterwordspacing

\bibitem{abdoulaye2021behavior}
A.~O. Ly and M.~Akhloufi, ``Learning to drive by imitation: An overview of deep
  behavior cloning methods,'' \emph{IEEE Transactions on Intelligent Vehicles},
  vol.~6, no.~2, pp. 195--209, 2021.

\bibitem{udrescu2020aifeynman}
S.-M. Udrescu and M.~Tegmark, ``Ai feynman: A physics-inspired method for
  symbolic regression,'' \emph{Science Advances}, vol.~6, no.~16, p. eaay2631,
  2020.

\bibitem{petersen2021deep}
\BIBentryALTinterwordspacing
B.~K. Petersen, M.~L. Larma, T.~N. Mundhenk, C.~P. Santiago, S.~K. Kim, and
  J.~T. Kim, ``Deep symbolic regression: Recovering mathematical expressions
  from data via risk-seeking policy gradients,'' in \emph{International
  Conference on Learning Representations}, 2021. [Online]. Available:
  \url{https://openreview.net/forum?id=m5Qsh0kBQG}
\BIBentrySTDinterwordspacing

\bibitem{landajuela2022a}
\BIBentryALTinterwordspacing
M.~Landajuela, C.~Lee, J.~Yang, R.~Glatt, C.~P. Santiago, I.~Aravena, T.~N.
  Mundhenk, G.~Mulcahy, and B.~K. Petersen, ``A unified framework for deep
  symbolic regression,'' in \emph{Advances in Neural Information Processing
  Systems}, A.~H. Oh, A.~Agarwal, D.~Belgrave, and K.~Cho, Eds., 2022.
  [Online]. Available: \url{https://openreview.net/forum?id=2FNnBhwJsHK}
\BIBentrySTDinterwordspacing

\bibitem{hao2020ns3ai}
\BIBentryALTinterwordspacing
H.~Yin, P.~Liu, K.~Liu, L.~Cao, L.~Zhang, Y.~Gao, and X.~Hei, ``Ns3-ai:
  Fostering artificial intelligence algorithms for networking research,'' in
  \emph{Proceedings of the 2020 Workshop on Ns-3}, ser. WNS3 2020.\hskip 1em
  plus 0.5em minus 0.4em\relax New York, NY, USA: Association for Computing
  Machinery, 2020, p. 57–64. [Online]. Available:
  \url{https://doi.org/10.1145/3389400.3389404}
\BIBentrySTDinterwordspacing

\bibitem{zhang2020reinforcement}
L.~Zhang, K.~Zhu, J.~Pan, H.~Shi, Y.~Jiang, and Y.~Cui, ``Reinforcement
  learning based congestion control in a real environment,'' in \emph{2020 29th
  International Conference on Computer Communications and Networks
  (ICCCN)}.\hskip 1em plus 0.5em minus 0.4em\relax IEEE, 2020, pp. 1--9.

\bibitem{tobin2017domain}
J.~{Tobin}, R.~{Fong}, A.~{Ray}, J.~{Schneider}, W.~{Zaremba}, and P.~{Abbeel},
  ``Domain randomization for transferring deep neural networks from simulation
  to the real world,'' in \emph{2017 IEEE/RSJ International Conference on
  Intelligent Robots and Systems (IROS)}, 2017, pp. 23--30.

\bibitem{martins2021policy}
J.~P. Martins, I.~Almeida, R.~Souza, and S.~Lins, ``Policy distillation for
  real-time inference in fronthaul congestion control,'' \emph{IEEE Access},
  vol.~9, pp. 154\,471--154\,483, 2021.

\end{thebibliography}
